\documentclass[aps,prl,preprint,groupedaddress]{revtex4-1}

\usepackage{color}
\begin{document}


\title{Extension of Nikiforov-Uvarov Method for the Solution of Heun Equation}


\author{H. Karayer\footnote{Author to whom correspondence should be addressed. Electronic mail: hale.karayer@gmail.com}}

\affiliation{Department of Physics, Faculty of Science, Ege University,35100 Bornova, Izmir, Turkey.}
\author{D. Demirhan}

\affiliation{Department of Physics, Faculty of Science, Ege University,35100 Bornova, Izmir, Turkey.}
\author{F. B\"uy\"ukk{\i}l{\i}\c{c}}
\affiliation{Department of Physics, Faculty of Science, Ege University,35100 Bornova, Izmir, Turkey.}

\date{\today}

\begin{abstract}
We report an alternative method to solve second order differential equations which have at most four singular points. This method is developed by changing the degrees of the polynomials in the basic equation of Nikiforov-Uvarov (NU) method. This is called extended NU method for this paper. The eigenvalue solutions of Heun equation and confluent Heun equation are obtained via extended NU method. Some quantum mechanical problems such as Coulomb problem on a 3-sphere, two Coulombically repelling electrons on a sphere and hyperbolic double-well potential are investigated by this method.   
\end{abstract}


\maketitle

\section{I. Introduction}

NU method is based on solving hypergeometric second order differential equations with special orthogonal functions \cite {szego}. Any second order differential equations which have three singular points can be converted to a hypergeometric type differential equation by a suitable change of variables. In spherical coordinates, for a given potential  Schr\"odinger equation is reduced to a generalized hypergeometric-type equation with appropriate coordinate transformation. Then this reduced equation can be solved systematically in order to obtain eigenvalues and eigenfunctions. The main equation of NU method is given in the following form \cite {Nikiforov};
\begin{equation}
\label{eq:nu}
\psi''(z)+\frac{\widetilde{\tau}(z)}{\sigma(z)}\psi'(z)+\frac{\widetilde{\sigma}(z)}{\sigma^{2}(z)}\psi(z)=0
\end{equation}
where $\widetilde{\tau}(z)$ is a polynomial of at most first-degree, $\sigma(z)$ and $\widetilde{\sigma}(z)$ are polynomials of at most second-degree and $\psi(z)$ is a function of hypergeometric-type.  The criteria which are related to degrees of polynomial coefficients constitute boundary conditions of the method.

The solution of Schr\"odinger equation is a fundamental problem in quantum mechanics for understanding the physical systems. Since, when the wave function $\psi(\vec{r},t)$ and energy eigenvalue are once obtained one has all possible information about the physical properties of the system \cite {berkdemir}. However the solution of the Schr\"odinger equation is widely complicated for a great number of potentials. To obtain an exact solution is only possible for a limited number of potentials such as harmonic oscillator potential, Coulomb potential, Kratzer potential, etc. \cite {berkdemir}. The standart solution method  is to expand the solution in a power series of independent variable and find the recursion relation for all the expansion coefficients \cite {flügge}. On the other hand, the power series solution method involves some detailed calculation techiques in order to reach the exact solution. Moreover in many cases usage of approximation methods becomes a necessity. However an approximated solution will roughly describe the system.

In present work the NU method is extended for second order differential equations that have at most four singularities. This is carried out by changing the degrees of polynomials in the main equation of NU method. In this extended method  $\widetilde{\tau}(z)$ will be a polynomial of at most second-degree, $\sigma(z)$ and $\widetilde{\sigma}(z)$ will be at most third and fourth-degrees respectively. Heun equation and confluent Heun equation can be given as examples for the equations that provide these boundary conditions of extended NU method. Heun differential equation and its confluent forms appear in several branches of physics. For a review see \cite{hortaçsu}. Also a particular solution of Heun equation has been studied in Ref.\cite{hale}.

This paper is organized as follows:

In second section the NU method is explained in details. 

In third section extended NU method is described and eigenvalue and eigenfunction solutions are given analytically.

In fourth and fifth sections eigenvalue and eigenfunction solutions of Heun equation and confluent Heun equation are derived via the extended NU method.

In sixth section we report some physical applications about the extended NU method.

\section{II. Nikiforov-Uvarov Method}
The foundation of NU method is to solve the second order hypergeometric type differential equations by means of special orthogonal functions. For a given potential, the radial Schr\"odinger equation is reduced to a generalized hypergeometric differential equation that is given by Eq.(\ref{eq:nu}). Several classes of special functions such as classical orthogonal polynomials, spherical harmonics, Bessel and hypergeometric functions are obtained from the solutions of this equation \cite{egrifes b}. By choosing an appropriate $\phi(z)$ and taking $\psi(z)=\phi(z)y(z)$, Eq.(\ref{eq:nu}) can be reduced to a more comprehensible form;
\begin{equation}
\label{eq:15}
\sigma(z)y''(z)+\tau(z)y'(z)+\lambda y(z)=0
\end{equation}
where 
\begin{equation}
\tau(z)=\widetilde{\tau}(z)+2\pi(z)
\end{equation}
This newly defined polynomial depends on polynomial $\pi(z)$ which is introduced as;
\begin{equation}
\label{eq:pi}
\pi(z)=\frac{\sigma'(z)-\widetilde{\tau}(z)}{2}\pm\sqrt{(\frac{\sigma'(z)-\widetilde{\tau}(z)}{2})^{2}-\widetilde{\sigma}(z)+k\sigma(z)}.
\end{equation}
In order to determine the polynomial $\pi(z)$, the parameter $k$ under the square root sign must be known and the expresion under the square root sign must be the square of a polynomial such that the requirement on $\pi(z)$ namely a polynomial of at most first degree has to be satisfied \cite {Nikiforov}. 

By using the property that, all the derivatives of hypergeometric functions are also of the hypergeometric type, the solutions of  Eq.(\ref{eq:15}) can be generalized. After this generalization the eigenvalue equation can be obtained as follows;
\begin{equation}
\lambda=\lambda_{n}=-n\tau'(z)-\frac{n(n-1)}{2}\sigma''(z)            (n=0,1,2,...).
\end{equation}
For the eigenfunction problem, there is a particular polynomial solution of the form $y(z)=y_{n}(z)$ which is given by Rodrigues formula \cite {Nikiforov};
\begin{equation}
y_{n}(z)=\frac{B_{n}}{\rho(z)}\frac{d^{n}}{dz^{n}}[\sigma^{n}(z)\rho(z)],
\end{equation}
where $B_{n}$ is a normalization constant and $\rho(z)$ is the weight function and this function satisfies the equation
$(\sigma(z)\rho(z))'=\tau(z)\rho(z)$.
The other function namely $\phi(z)$ in the $\psi(z)=\phi(z)y(z)$ transformation is defined as a logarithmic derivative;
\begin{equation}
\frac{\phi'(z)}{\phi(z)}=\frac{\pi(z)}{\sigma(z)}. 
\end{equation}
After determination of all the polynomial parameters, the eigenvalue and the eigenfunction solutions of Schr\"odinger equation can be obtained completely in a systematic way by using NU method.

\section{III. Extended Nikiforov-Uvarov Method}
This method is improved for solving any second order differential equation which has at most four singular points. The improvement is achieved by changing the boundary conditions of NU method. An equation provided that the boundary conditions of extended NU method are satisfied, is reduced by an appropriate coordinate transformation into the following form;
\begin{equation}
\label{eq:ex_nu}
\psi''(z)+\frac{\widetilde{\tau_{e}}(z)}{\sigma_{e}(z)}\psi'(z)+\frac{\widetilde{\sigma_{e}}(z)}{\sigma_{e}^{2}(z)}\psi(z)=0
\end{equation}
where $\widetilde{\tau}_{e}(z)$, $\sigma_{e}(z)$ and $\widetilde{\sigma}_{e}(z)$ are polynomials of at most second, third and fourth-degrees respectively. By choosing a suitable function $ \phi_{e}(z) $ and taking;
\begin{equation}
\label{eq:ex_psi}
\psi(z)=\phi_{e}(z)y(z)
\end{equation}
Eq.(\ref{eq:ex_nu}) is converted to the following form:
\begin{equation}
\label{eq:ex_y}
y''(z)+\bigg(2\frac{\phi'_{e}(z)}{\phi_{e}(z)}+\frac{\widetilde{\tau_{e}}(z)}{\sigma_{e}(z)}\bigg)y'(z)+\bigg(\frac{\phi''_{e}(z)}{\phi_{e}(z)}+\frac{\phi'_{e}(z)}{\phi_{e}(z)}\frac{\widetilde{\tau_{e}}(z)}{\sigma_{e}(z)}+\frac{\widetilde{\sigma_{e}}(z)}{\sigma_{e}^{2}(z)}\bigg)y(z)=0.
\end{equation}
To obtain a simplified form of Eq.(\ref{eq:ex_y}) the coefficients of $ y(z) $ and $ y'(z) $ are arranged with newly defined polynomials. The coefficient of $ y'(z) $ is taken in the form  $ {\tau_{e}(z)}/{\sigma_{e}(z)} $; 
\begin{equation}
\label{arg1}
2\frac{\phi'_{e}(z)}{\phi_{e}(z)}+\frac{\widetilde{\tau_{e}}(z)}{\sigma_{e}(z)}=\frac{\tau_{e}(z)}{\sigma_{e}(z)}.
\end{equation}
where $ \tau_{e}(z) $ is a polynomial of at most second degree. The most regular form of this equation is obtained as follows;
\begin{equation}
\label{phi}
\frac{\phi'_{e}(z)}{\phi_{e}(z)}=\frac{\pi_{e}(z)}{\sigma_{e}(z)}
\end{equation}
where $\pi_{e}(z)$, provided that
\begin{equation}
\label{eq:ex_tau}
\tau_{e}(z)=\widetilde{\tau_{e}}(z)+2\pi_{e}(z),
\end{equation}
is a polynomial of at most second-degree. In addition for the coefficient of $y(z)$, Eq.(\ref{eq:ex_y}) is also rearranged as follows;
\begin{equation}
\label{arg2}
\frac{\phi''_{e}(z)}{\phi_{e}(z)}+\frac{\phi'_{e}(z)}{\phi_{e}(z)}\frac{\widetilde{\tau_{e}}(z)}{\sigma_{e}(z)}+\frac{\widetilde{\sigma_{e}}(z)}{\sigma_{e}^{2}(z)}=\frac{\bar{\sigma_{e}}(z)}{\sigma_{e}^{2}(z)}.
\end{equation}
where $\bar{\sigma_{e}}(z)$ is a polynomial of at most fourth degree and it is defined as;
\begin{equation}
\label{eq:ex_sigmabar}
\bar{\sigma_{e}}(z)=\widetilde{\sigma_{e}}(z)+\pi_{e}^{2}(z)+\pi_{e}(z)(\widetilde{\tau_{e}}(z)-\sigma_{e}'(z))+\pi_{e}'(z)\sigma_{e}(z).
\end{equation}
After substituting the right-hand sides of  Eq.(\ref{arg1}) and Eq.(\ref{arg2}) into Eq.(\ref{eq:ex_y}) the following equation can be obtained:
\begin{equation}
\label{eq:ex_d}
y''(z)+\frac{\tau_{e}(z)}{\sigma_{e}(z)}y'(z)+\frac{\bar{\sigma_{e}}(z)}{\sigma^{2}_{e}(z)}y(z)=0.
\end{equation}
If the polynomial $\bar{\sigma_{e}}(z)$ is divisible by $\sigma_{e}(z)$, i.e.,  
\begin{equation}
{\bar{\sigma_{e}}(z)}/{\sigma_{e}(z)}=h(z)
\end{equation}
where $ h(z) $ is a polynomial of degree at most one, Eq.(\ref{eq:ex_d}) is reduced to the following form;
\begin{equation}
\label{asl}
\sigma_{e}(z)y''(z)+\tau_{e}(z)y'(z)+h(z)y(z)=0.
\end{equation}
If the definition $ {\bar{\sigma_{e}}(z)}/{\sigma_{e}(z)}=h(z) $ is also used in Eq.(\ref{eq:ex_sigmabar}), the second order homogeneous equation is obtained as 
\begin{equation}
\label{kök}
\pi_{e}^{2}(z)+\pi_{e}(z)(\widetilde{\tau_{e}}(z)-\sigma_{e}'(z))+\widetilde{\sigma_{e}}(z)-(h(z)-\pi_{e}'(z))\sigma_{e}(z)=0,
\end{equation}
where
\begin{equation}
\label{eq:ex_h(z)}
h(z)-\pi_{e}'(z)=g(z).
\end{equation}
Using the roots of Eq.(\ref{kök}), $ \pi_{e}(z) $ can be obtained as follows;
\begin{equation}
\label{eq:ex_pi}
\pi_{e}(z)=\frac{\sigma_{e}'(z)-\widetilde{\tau_{e}}(z)}{2}\pm\sqrt{(\frac{\sigma_{e}'(z)-\widetilde{\tau_{e}}(z)}{2})^{2}-\widetilde{\sigma_{e}}(z)+g(z)\sigma_{e}(z)}.
\end{equation}
To determine all of possible solutions of the polynomial $ \pi_{e}(z) $,  the polynomial $ g(z) $ under the square root sign must be known explicitly. Since $ \pi_{e}(z) $ is a polynomial of at most second-degree, the expression under the square root sign must be square of a polynomial of at most second-degree. Therefore, the polynomials  $ g(z) $ which satisfy this condition are chosen and after determination of $ g(z) $, the polynomial
 $ \pi_{e}(z) $ can be obtained from Eq.(\ref{eq:ex_pi}). Then $ \tau_{e}(z) $ and $ h(z) $ can be found from 
Eq.(\ref{eq:ex_tau}) and Eq.(\ref{eq:ex_h(z)}) respectively.

In order to generalize the solutions of Eq.(\ref{asl}), this equation is differentiated once and a compherensible form is obtained:
\begin{equation}
\label{hip}
\sigma_{e}(z)y^{(3)}(z)+(\tau_{e}(z)+\sigma_{e}'(z))y''(z)+(\tau_{e}'(z)+h(z))y'(z)+h'(z)y(z)=0.
\end{equation}
This equation is a third-order homogeneous differential equation with polynomial coefficients of degree not exceeding the corresponding order of differentiation. Since all derivatives of Eq.(\ref{hip}) have the same form, it can be differentiated $n$ times by using the new representation $y^{(n)}(z)=v_{n}(z)$; 
\begin{eqnarray}
\label{poly}
\sigma_{e}(z)v_{n}^{(3)}(z)+(\tau_{e}(z)+(n+1)\sigma_{e}'(z))v_{n}''(z)+\nonumber \\ ((n+1)\tau_{e}'(z)+\frac{n(n+1)}{2}\sigma_{e}''(z)+h(z))v_{n}'(z)+ \nonumber \\
((n+1)h'(z)+\frac{n(n+1)}{2}\tau_{e}''(z)+\frac{n(n+1)(n-1)}{6}\sigma_{e}^{(3)}(z))v_{n}(z)=0
\end{eqnarray}
When the coefficient of $ v_{n}(z) $ is equal to zero, polynomial $ h_{n}(z) $ is defined as
\begin{equation}
\label{hn}
h_{n}(z)=-\frac{n}{2}\tau_{e}'(z)-\frac{n(n-1)}{6}\sigma_{e}''(z)+C_{n},
\end{equation}
where $C_{n}$ is an integration constant. Therefore Eq.(\ref{poly}) is reduced into the following form:
\begin{eqnarray}
\label{polyson}
\sigma_{e}(z)v_{n}^{(3)}(z)+\bigg(\tau_{e}(z)+(n+1)\sigma_{e}'(z)\bigg)v_{n}''(z)+\nonumber \\
\bigg((\frac{n}{2}+1)\tau_{e}'(z)+\frac{n(n+2)}{3}\sigma_{e}''(z)+C_{n}\bigg)v_{n}'(z) =0.
\end{eqnarray}
Eq.(\ref{polyson}) has a particular solution of the form $ y(z)=y_{n}(z) $ which is a polynomial of degree $n$. To obtain the eigenvalue solution, the relationship between $h(z)$ and $h_{n}(z)$ must be set up by means of Eq.(\ref{eq:ex_h(z)}) and Eq.(\ref{hn}).

\section{IV. Solution of Heun Equation by Extended Nikiforov-Uvarov Method }
Heun equation in which Gauss hypergeometric, confluent hypergeometric, Mathieu, Ince, Legendre, Laguerre, Bessel functions are involved, is a general second order linear differential equation:
\begin{equation}
\label{eq:heun}
\frac{d^2w}{dz^2}+[\frac{\gamma}{z}+\frac{\delta}{z-1}+\frac{\epsilon}{z-a}]\frac{dw}{dz}+\frac{\alpha\beta z-q}{z(z-1)(z-a)}w=0.
\end{equation}
where $\gamma$, $\delta$, $\epsilon$, $\alpha$ and $\beta$ are constant parameters which satisfy the Fuchsian condition; $\epsilon=\alpha+\beta-\gamma-\delta+1$. The complex number $q$ is called accessory parameter. Eq.(\ref{eq:heun}) has four regular singular points at $z=0$, $z=1$, $z=\infty$ and $z=a$ \cite{hounkonnou}. The most widely studied solutions of Heun's equation literature are Heun polynomials which could be classified into eight classes. These solutions are written in the form;
\begin{equation}
Hp(z)=z^{\sigma_{1}}(z-1)^{\sigma_{2}}(z-a)^{\sigma_{3}}p(z)
\end{equation}
where $ p(z) $ is a polynomial and $ \sigma_{1}, \sigma_{2}, \sigma_{3} $  are each one of the two possible exponents at the corresponding singularity. The classification of Heun polynomials is demonstrated in a table that is given as \cite{heun};\\

\begin{tabular}{cccccc}
\hline
Class & $ \sigma_{1}$ & $ \sigma_{2}$ & $ \sigma_{3} $&$ \alpha $&$ \beta $ \\
\hline
I& $ 0 $ &$ 0 $ & $ 0 $&$ -n $ &$ \epsilon+\gamma+\delta+n-1 $ \\
II &$ 1-\gamma $ &$ 0 $ &$ 0 $ & $ \gamma-n-1 $& $ n+\delta+\epsilon $\\
III & $ 0 $&$ 1-\delta $ & $ 0$& $\delta-n-1  $ & $ \epsilon+\gamma+n $\\
IV &$ 1-\gamma $ &$ 1-\delta $ & $ 0 $& $ \delta+\gamma-n-2 $&$ \epsilon+n+1 $ \\
V & $ 0 $&$ 0 $ &$ 1-\epsilon $ & $\epsilon-n-1  $ & $ \delta+\gamma+n $\\
VI &$ 1-\gamma $ &$ 0 $ &$ 1-\epsilon $ & $ \epsilon+\gamma-n-2 $&$ \delta+n+1 $ \\
VII &$ 0 $ & $ 1-\delta $&$ 1-\epsilon $ & $ \delta+\epsilon-n-2 $& $ \gamma+n+1 $\\
VIII &$ 1-\gamma $ &$ 1-\delta $ &$ 1-\epsilon $ & $ \epsilon+\gamma+\delta-n-3 $&$ n+2 $ \\
\hline
\end{tabular}
\begin{center}
Table-I: \textit{Classification of Heun Polynomials}
\end{center}
Heun equation yields the boundary conditions of extended NU method. Comparing Eq.(\ref{eq:heun}) and Eq.(\ref{eq:ex_nu}), the relationship between polynomials in the extended NU method and the parameters in Heun equation is set up as follows;
\begin{eqnarray}
\widetilde{\tau}_{e}(z)&=&\gamma(z-1)(z-a)+\delta z(z-a)+\epsilon z(z-1)\\
\sigma_{e}(z)&=&z(z-1)(z-a)\\
\widetilde{\sigma}_{e}(z)&=&(\alpha\beta z-q)z(z-1)(z-a).
\end{eqnarray}
Substituting these polynomials into Eq.(\ref{eq:ex_pi}) $\pi_{e}(z)$ polynomial can be obtained as;

\begin{eqnarray}
\pi_{e}(z)=\frac{(1-\gamma)(z-1)(z-a)+(1-\delta)z(z-a)+(1-\epsilon)z(z-1)}{2}\pm   \nonumber\\
\frac{1}{2}\bigg\{ [(1-\gamma)(z-1)(z-a)+(1-\delta)z(z-a)+(1-\epsilon)z(z-1)]^2-
 \nonumber\\
4(\alpha\beta z-q)z(z-1)(z-a)+4g(z)z(z-1)(z-a)\bigg\}^\frac{1}{2} \nonumber
\end{eqnarray}

Since $\pi(z)$ is defined as a second order polynomial, the expression under the square root sign has to be the square of a second order polynomial. Hence all of the possible solutions of $\pi(z)$ corresponding to certain values of polynomials $g(z)$ can be listed as follows:

\begin{eqnarray}
\label{p11}
\pi_{e1}(z)=(1-\gamma)(z-1)(z-a)+(1-\delta)z(z-a)+(1-\epsilon)z(z-1) \\
\label{p12}
\pi_{e2}(z)=0
\end{eqnarray}
for $g_{1}(z)=\alpha\beta z-q$

\begin{eqnarray}
\label{p21}
\pi_{e3}(z)=(1-\gamma)(z-1)(z-a) \\
\label{p22}
\pi_{e4}(z)=(1-\delta)z(z-a)+(1-\epsilon)z(z-1)
\end{eqnarray}
for $g_{2}(z)=\alpha\beta z-q-(1-\gamma)[(1-\delta)(z-a)+(1-\epsilon)(z-1)]$

\begin{eqnarray}
\label{p31}
\pi_{e5}(z)=(1-\gamma)(z-1)(z-a)+(1-\delta)z(z-a) \\
\label{p32}
\pi_{e6}(z)=(1-\epsilon)z(z-1)
\end{eqnarray}
for $g_{3}(z)=\alpha\beta z-q-(1-\epsilon)[(1-\gamma)(z-1)+(1-\delta)z]$

\begin{eqnarray}
\label{p41}
\pi_{e7}(z)=(1-\gamma)(z-1)(z-a)+(1-\epsilon)z(z-1) \\
\label{p42}
\pi_{e8}(z)=(1-\delta)z(z-a)
\end{eqnarray}
for $g_{4}(z)=\alpha\beta z-q-(1-\delta)[(1-\gamma)(z-a)+(1-\epsilon)z]$\\

Starting from $ \pi_{e1}(z) $ in Eq.(\ref{p11}) the analytical solution can be written in details as follows: $ h(z) $ can be obtained from Eq.(\ref{eq:ex_h(z)});
\begin{equation}
\label{h1}
h(z)=\alpha\beta z-q+2z[3-(\epsilon+\gamma+\delta)]-(1-\gamma)(a+1)-(1-\delta)a-(1-\epsilon),
\end{equation}
Then from Eq.(\ref{hn}) polynomial $ h_{n}(z) $ is achieved as; 
\begin{eqnarray}
\label{h2}
h_{n}(z)=-n[n+5-(\epsilon+\gamma+\delta)]z+\nonumber \\\frac{n}{2}[(a+1)(2-\gamma)+a(2-\delta)+(2-\epsilon)
+\frac{2}{3}(n-1)(a+1)]+C_{n1},
\end{eqnarray}
By making use of the equality $h(z)= h_{n}(z) $ the eigenvalue equation and $ q $ accessory parameter could be calculated and are given by;

\begin{eqnarray}
\alpha\beta=(n+2)(\epsilon+\gamma+\delta-n-3)\\
-q=\frac{n}{2}[(a+1)(2-\gamma)+a(2-\delta)+(2-\epsilon)+\frac{2}{3}(n-1)(a+1)]+\nonumber \\
(1-\gamma)(a+1)+(1-\delta)a+(1-\epsilon)+C_{n1}.
\end{eqnarray}
The eigenfunction solution can be also acquired as follows; Firstly the function $ \phi_{e}(z) $ is calculated from Eq.(\ref{phi}):
\begin{equation}
\phi_{e}(z)=z^{1-\gamma}(z-1)^{1-\delta}(z-a)^{1-\epsilon}.
\end{equation}
Then a polynomial solution of degree $n$ of Eq.(\ref{polyson}) is obtained and demonstrated by $ p_{1}(z)  $. Substituting these into Eq.(\ref{eq:ex_psi}), the eigenfunction can be achieved analytically;
\begin{equation}
\label{psi1}
\psi_{1}(z)=\phi_{e}(z)y(z)=z^{1-\gamma}(z-1)^{1-\delta}(z-a)^{1-\epsilon}p_{1}(z).
\end{equation}
This solution corresponds to the class VIII Heun polynomial \cite{heun}. Since the above mentioned procedure is valid for all of the polynomials $ \pi_{e}(z) $ given in Eq.(\ref{p12}-\ref{p42}) in the eigenvalue and eigenfunction solutions, one can directly write down these solutions:\\

For Eq.(\ref{p12});
\begin{eqnarray}
\label{heunI}
\alpha\beta=-n(\epsilon+\gamma+\delta+n-1)\\
\label{q1}
-q=\frac{n}{2}[(a+1)\gamma+a\delta+\epsilon+\frac{2}{3}(n-1)(a+1)]+C_{n2}
\end{eqnarray}
\begin{equation}
\label{psi2}
\psi_{2}(z)=z^{0}(z-1)^{0}(z-a)^{0}p_{2}(z)
\end{equation}
This solution corresponds to class I Heun polynomial \cite{heun}.\\

For Eq.(\ref{p21});
\begin{eqnarray}
\label{heunII}
\alpha\beta=(n+\delta+\epsilon)(\gamma-n-1) \\
-q=\frac{n}{2}[(a+1)(2-\gamma)+a\delta+\epsilon+\frac{2}{3}(n-1)(a+1)]+(1-\gamma)(a\delta+\epsilon)+C_{n3}
\end{eqnarray}

\begin{equation}
\label{psi3}
\psi_{3}(z)=z^{1-\gamma}(z-1)^{0}(z-a)^{0}p_{3}(z)
\end{equation}
which corresponds to class II Heun polynomial \cite{heun}.\\

For Eq.(\ref{p22});
\begin{eqnarray}
\alpha\beta=(\delta+\epsilon-n-2)(\gamma+n+1) \\
-q=\frac{n}{2}[(a+1)\gamma+a(2-\delta)+(2-\epsilon)+\frac{2}{3}(n-1)(a+1)]+ \nonumber \\
\gamma[a(1-\delta)-(1-\epsilon)]+C_{n4}
\end{eqnarray}

\begin{equation}
\label{psi4}
\psi_{4}(z)=z^{0}(z-1)^{1-\delta}(z-a)^{1-\epsilon}p_{4}(z)
\end{equation}
which is the corresponding class VII Heun polynomial \cite{heun}.\\

For Eq.(\ref{p31});
\begin{eqnarray}
\alpha\beta=(\delta+\gamma-n-2)(\epsilon+n+1) \\
-q=\frac{n}{2}[(a+1)(2-\gamma)+a(2-\delta)+\epsilon+
\frac{2}{3}(n-1)(a+1)]+\nonumber \\(1-\gamma)(a+\epsilon)+(1-\delta)a+C_{n5}
\end{eqnarray}

\begin{equation}
\label{psi5}
\psi_{5}(z)=z^{1-\gamma}(z-1)^{1-\delta}(z-a)^{0}p_{5}(z)
\end{equation}
namely the corresponding class IV Heun polynomial \cite{heun}.\\

For Eq.(\ref{p32});
\begin{eqnarray}
\alpha\beta=(\delta+\gamma+n)(\epsilon-n-1) \\
-q=\frac{n}{2}[(a+1)\gamma+a\delta+\epsilon-2+\frac{2}{3}(n-1)(a+1)]-(2-\gamma)(1-\epsilon)+C_{n6}
\end{eqnarray}

\begin{equation}
\label{psi6}
\psi_{6}(z)=z^{0}(z-1)^{0}(z-a)^{1-\epsilon}p_{6}(z)
\end{equation}
Eq.(\ref{psi6}) is the corresponding class V Heun polynomial \cite{heun}.\\

For Eq.(\ref{p41});
\begin{eqnarray}
\alpha\beta=(\epsilon+\gamma-n-2)(\delta+n+1) \\
-q=\frac{n}{2}[(a+1)(2-\gamma)+a\delta+(2-\epsilon)+\frac{2}{3}(n-1)(a+1)]+\nonumber \\
(1-\gamma)(a+\delta)+(1-\epsilon)a+C_{n7}
\end{eqnarray}

\begin{equation}
\label{psi7}
\psi_{7}(z)=z^{1-\gamma}(z-1)^{0}(z-a)^{1-\epsilon}p_{7}(z)
\end{equation}
This solution is the corresponding class VI Heun polynomial \cite{heun}.\\

For Eq.(\ref{p42});
\begin{eqnarray}
\alpha\beta=(\epsilon+\gamma+n)(\delta-n-1) \\
-q=\frac{n}{2}[(a+1)\gamma+a(2-\delta)+\epsilon+\frac{2}{3}(n-1)(a+1)]-\nonumber \\
(2-\gamma)(1-\delta)a+C_{n8}
\end{eqnarray}

\begin{equation}
\label{psi8}
\psi_{8}(z)=z^{0}(z-1)^{1-\delta}(z-a)^{0}p_{8}(z)
\end{equation}
Eq.(\ref{psi8}) gives the corresponding class III Heun polynomial \cite{heun}.\\

For the existence of polynomial solutions of Heun equation, one of the parameters $ \alpha $ or $ \beta $ must be chosen conveniently. For instance, for the class VIII Heun polynomial these parameters are chosen to be $ \alpha=\epsilon+\gamma+\delta-n-3 $ and $ \beta = n+2$ \cite{heun}. In the solution which is obtained by extended NU method, the parameters $ \alpha $ and $ \beta $ can be determined exactly. Also the accessory parameter $q$ is specified exactly up to a constant of integration. \\

\section{V. Solution of Confluent Heun Equation by Extended Nikiforov-Uvarov Method }
The 'confluent Heun equation' is derived from the Heun equation by the coincidence of singularities at $z=a$ and $z=\infty$. The general confleunce process is discussed in details in Ince \cite{Ince}. Also derivation of the confluent Heun equation (CHE) is given as an illustration in Ref. \cite{heun}. The most general form of CHE is expressed as;
\begin{eqnarray}
\label{che}
\psi''(z)+\bigg(\sum_{i=1}^{2}\frac{A_{i}}{z-z_{i}}+\sum_{j=1}^{2}\frac{E_{3j}}{(z-z_{3})^{j}}\bigg)\psi'(z)+\nonumber \\
\bigg(\sum_{i=1}^{2}\frac{C_{i}}{z-z_{i}}+\sum_{i=1}^{2}\frac{B_{i}}{(z-z_{i})^{2}}+\sum_{j=1}^{4}\frac{D
_{3j}}{(z-z_{3})^{j}}\bigg)\psi(z)=0
\end{eqnarray}
where the location of singular points and coefficients are arbitrary. The simplest uniform shape of Eq.(\ref{che}) can be written as \cite{ciftci};
\begin{equation}
\label{che1}
y''(z)+\bigg(\alpha +\frac{\beta+1}{z}+\frac{\gamma+1}{z-1}\bigg)y'(z)+\bigg(\frac{\mu}{z}+\frac{\nu}{z-1}\bigg)y(z)=0.
\end{equation} 
This equation has regular singularities at $z=0$ and $z=1$, and an irregular singularity at  $z=\infty$. The solution of Eq.(\ref{che1}) in the vicinity of the regular singular point $z=0$ is given by confluent Heun function \cite {fiziev}
\begin{equation}
H_{C}(\alpha, \beta, \gamma, \delta, \eta, z)=\sum\limits_{n=0}^{\infty}v_{n}(\alpha, \beta, \gamma, \delta, \eta, z)z^{n},
\end{equation}
where
\begin{eqnarray}
\delta = \mu+\nu-\frac{\alpha}{2}(\beta+\gamma+2) \nonumber \\
\eta=\frac{\alpha}{2}(\beta+1)-\mu-\frac{1}{2}(\beta+\gamma+\beta\gamma)
\end{eqnarray}
and the coefficients $v_{n}$ are given by three-term recurrence relation \cite{heun}. This solution can be reduced to a confluent Heun polynomial of degree $n$  provided that the following conditions are satisfied:
\begin{equation}
\mu +\nu=-n\alpha
\end{equation} 
together with the vanishing of the $(n+1)$x$(n+1)$ tridiagonal determinant given by \cite{downing};\\

\begingroup\makeatletter\def\f@size{8}\check@mathfonts
{\begin{tabular}{|ccccccccc|}
 $ \mu-q_{1} $ &  $ ( 1+\beta)$  & $ 0 $   & $ . $   &  $ . $  &  $ . $  &  $ 0 $  &  $ 0 $  &  $ 0 $  \\ 
 $ n\alpha $ &  $ \mu-q_{2}+\alpha $  & $ 2( 2+\beta) $   & $ . $   &  $ . $  &  $ . $  &  $ 0 $  &  $ 0 $  &  $ 0 $   \\ 
 $ 0 $ &  $  (n-1)\alpha $  & $ \mu-q_{3}+2\alpha $   & $ . $   &  $ . $  &  $ . $  &  $ 0 $  &  $ 0 $  &  $ 0 $  \\ 
 $ . $ &  $ . $  & $ . $   & $ . $   &  $  $  &  $  $  &  $ . $  &  $ . $  &  $ . $  \\ 
 $ . $ &  $ . $  & $ . $   & $  $   &  $ . $  &  $  $  &  $ . $  &  $ . $  &  $ . $  \\ 
 $ . $ &  $ . $  & $ . $   & $  $   &  $  $  &  $ . $  &  $ . $  &  $ . $  &  $ . $  \\ 
 $ 0 $ &  $ 0 $  & $ 0 $   & $ . $   &  $ . $  &  $ . $  &  $ \mu-q_{n-1}+(n-2)\alpha $  &  $ (n-1)(n-1+\beta) $  &  $ 0 $  \\ 
 $ 0 $ &  $ 0 $  & $ 0 $   & $ . $   &  $ . $  &  $ . $  &  $ 2\alpha $  &  $  \mu-q_{n}+(n-1)\alpha $  &  $ n(n+\beta) $  \\ 
 $ 0 $ &  $ 0 $  & $ 0 $   & $ . $   &  $ . $  &  $ . $  &  $ 0 $  &  $ \alpha $  &  $  \mu-q_{n+1}+n\alpha $  \\ 
\end{tabular} }=0.
\endgroup \\

The CHE which is given by Eq.(\ref{che1}), could analytically be solved by extended NU method as well. The solution process begins by the comparision of the boundary conditions of extended NU method with the degrees of polynomial coefficients in CHE. After this comparision the polynomials in the extended NU method can be associated with the parameters in the CHE as follows:
\begin{eqnarray}
\widetilde{\tau}_{e}(z)&=&\alpha z(z-1)+(\beta +1) (z-1)+(\gamma +1) z\\
\sigma_{e}(z)&=&z(z-1)\\
\widetilde{\sigma}_{e}(z)&=&[(\mu +\nu) z-\mu]z(z-1).
\end{eqnarray}
Subsituting these polynomials into Eq.(\ref{eq:ex_pi}), the polynomial $\pi_{e}(z)$ can be obtained:
\begin{eqnarray}
\pi_{e}(z)=\frac{-\alpha z(z-1)-\beta (z-1)-\gamma z}{2}\pm \nonumber \\
\frac{1}{2}\sqrt{\big[-\alpha z(z-1)-\beta (z-1)-\gamma z\big]^{2}-
4[(\mu +\nu)z-\mu-g(z)]z(z-1)}.
\end{eqnarray}
Since $\pi_{e}(z)$ is a second degree polynomial, the expression under the square root must be a second degree polynomial. For this purpose polynomials $g(z)$ are choosen as follows;

\begin{eqnarray}
\label{cp11}
\pi_{e1}(z)= -\alpha z(z-1)-\beta (z-1)-\gamma z\\
\label{cp12}
\pi_{e2}(z)=0
\end{eqnarray}
for $g_{1}(z)=(\mu +\nu)z-\mu$

\begin{eqnarray}
\label{cp21}
\pi_{e3}(z)= -\gamma z\\
\label{cp22}
\pi_{e4}(z)=-\alpha z(z-1)-\beta (z-1)
\end{eqnarray}
for $g_{2}(z)=(\mu +\nu -\gamma \alpha)z-\mu- \gamma \beta$

\begin{eqnarray}
\label{cp31}
\pi_{e5}(z)= -\beta (z-1)-\gamma z\\
\label{cp32}
\pi_{e6}(z)=-\alpha z(z-1)
\end{eqnarray}
for $g_{3}(z)=(\mu +\nu -\alpha \beta-\gamma \alpha)z-\mu- \alpha \beta$

\begin{eqnarray}
\label{cp41}
\pi_{e7}(z)= -\beta (z-1)\\
\label{cp42}
\pi_{e8}(z)=-\alpha z(z-1)-\gamma z
\end{eqnarray}
for $g_{4}(z)=(\mu +\nu -\alpha \beta)z-\mu+\alpha \beta-\beta \gamma$\\

For $ \pi_{e1}(z) $ in Eq.(\ref{cp11}), the analytical solution is given as follows: polynomial $ h(z) $ can be obtained from Eq.(\ref{eq:ex_h(z)});
\begin{equation}
\label{a}
h(z)=(\mu +\nu-2\alpha)z-\mu+\alpha-\beta-\gamma.
\end{equation}
polynomial $h_{n}(z)$ is achieved from Eq.(\ref{poly}) when the coefficient of $ v_{n}(z) $ is equal to zero as mentioned in Section.3. But in the CHE problem $\sigma(z)$ is a second degree polynomial such that $\sigma^{(3)}(z)$ in the coefficient of $ v_{n}(z) $ is equal to zero. Then $ h_{n}(z) $ is:
\begin{equation}
\label{b}
h_{n}(z)=-\frac{n}{2}\tau'(z)+C_{n}=-\frac{n}{2}(2\alpha z+2-\alpha+\beta+\gamma)+C_{n}.
\end{equation}
By equating the right hand sides of Eq.(\ref{a}) and Eq.(\ref{b}), eigenvalue equation and parameter $\mu$ (with a difference up to an integration constant) are achieved;
\begin{eqnarray}
\label{c}
\mu+\nu-2\alpha=n\alpha \\
-\mu=-\frac{n}{2}(2+\alpha+\beta-\gamma)-\alpha+\beta+\gamma+C_{n1}.
\end{eqnarray}
For the eigenfunction solution of CHE, $ \phi_{e}(z) $ and the polynomial part of the solution can be specified from Eq.(\ref{phi}) and Eq.(\ref{poly}) respectively. The function $ \phi_{e}(z) $ is;
\begin{equation}
\phi_{e}(z)=e^{-\alpha z} z^{-\beta} (z-1)^{-\gamma}.
\end{equation}
If the polynomial solution is indicated by $ p_{1}(z) $, by using Eq.(\ref{eq:ex_psi}) the eigenfunction solution can be obtained completely;
\begin{equation}
\label{d}
\psi_{1}(z)=\phi_{e}(z)y(z)=e^{-\alpha z} z^{-\beta} (z-1)^{-\gamma}p_{1}(z).
\end{equation}
Since the solution process given for $ \pi_{e1}(z) $ will be same for all of the polynomials $ \pi_{e}(z) $ given in Eq.(\ref{cp12}-\ref{cp42}), the results can be written down directly:

For Eq.(\ref{cp12});
\begin{eqnarray}
\label{sym1}
\mu+\nu=-n\alpha \\
\label{mu1}
\mu=\frac{n}{2}(2-\alpha+\beta+\gamma)+C_{n2} \\
\psi_{2}(z)=p_{2}(z).
\end{eqnarray}

For Eq.(\ref{cp21});
\begin{eqnarray}
\label{asym1}
\mu+\nu-\gamma \alpha=-n\alpha \\
-\mu=-\frac{n}{2}(2-\alpha+\beta-\gamma)+\gamma(1+\beta)+C_{n3} \\
\psi_{3}(z)=(z-1)^{-\gamma}p_{3}(z).
\end{eqnarray}

For Eq.(\ref{cp22});
\begin{eqnarray}
\mu+\nu-\gamma \alpha-2\alpha=n\alpha \\
-\mu=-\frac{n}{2}(2+\alpha-\beta+\gamma)-\alpha+\beta+\gamma\beta+C_{n4} \\
\psi_{4}(z)=e^{-\alpha z} z^{-\beta}p_{4}(z).
\end{eqnarray}

For Eq.(\ref{cp31});
\begin{eqnarray}
\label{asym2}
\mu+\nu-\gamma \alpha-\beta\alpha=-n\alpha \\
-\mu=-\frac{n}{2}(2+\alpha-\beta+\gamma)-\alpha+\beta+\gamma\beta+C_{n5} \\
\psi_{5}(z)=z^{-\beta}(z-1)^{-\gamma}p_{5}(z).
\end{eqnarray}

For Eq.(\ref{cp32});
\begin{eqnarray}
\mu+\nu-\gamma \alpha-\beta\alpha-2\alpha=n\alpha \\
-\mu=-\frac{n}{2}(2+\alpha+\beta+\gamma)-\alpha-\alpha\beta+C_{n6} \\
\psi_{6}(z)=e^{-\alpha z}p_{6}(z).
\end{eqnarray}

For Eq.(\ref{cp41});
\begin{eqnarray}
\label{sym2}
\mu+\nu-\beta\alpha=n\alpha \\
-\mu=-\frac{n}{2}(2+\alpha-\beta+\gamma)-\alpha\beta+\beta(\gamma+1)+C_{n7} \\
\psi_{7}(z)=z^{-\beta}p_{7}(z).
\end{eqnarray}

For Eq.(\ref{cp42});
\begin{eqnarray}
\mu+\nu-\beta\alpha-2\alpha=n\alpha \\
-\mu=-\frac{n}{2}(2+\alpha+\beta-\gamma)-\alpha-\alpha\beta+\gamma(1+\beta)+C_{n8} \\
\psi_{8}(z)=e^{-\alpha z}(z-1)^{-\gamma}p_{8}(z).
\end{eqnarray}

\section{VI. Applications}
\subsection{1. The Coulomb problem on a 3-sphere}
In Ref.\cite{bellucci}, an analytical solution of the Schr\"odinger equation for the Coulomb problem on 3-sphere in generalized parabolic coordinates is given. After using the method of seperation of variables for the wave function in the Schr\"odinger equation, they obtain the Heun equation in the following form;
\begin{equation}
\label{hlc}
Hl''+(\frac{\Gamma}{z}+\frac{\Delta}{z-1}+\frac{\epsilon}{z+1})Hl'+\frac{abz-q}{z(z-1)(z+1)}Hl=0.
\end{equation}
where
\begin{eqnarray}
\label{qs}
\Gamma=1-\sqrt{1+E+i\gamma}, \nonumber \\
\Delta_{m}=\epsilon_{m}=\mid m \mid +1 \nonumber \\
a=1+\mid m \mid +\frac{\sqrt{1+E-i\gamma}-\sqrt{1+E+i\gamma}}{2},\nonumber \\
b=1+\mid m \mid -\frac{\sqrt{1+E-i\gamma}+\sqrt{1+E+i\gamma}}{2}.
\end{eqnarray}
In the above mentioned reference the eigenvalue solution of Eq.(\ref{hlc}) is obtained in the form;
\begin{equation}
E_{n}=(n+\mid m \mid)(n+\mid m \mid+2)-\frac{\gamma^{2}}{4(n+\mid m \mid+1)^{2}}.
\end{equation}

If the expressions given in Eq(\ref{qs}) are substituted in Eq.(\ref{heunI}) and Eq.(\ref{heunII}) the same solution can be obtained by extended NU method;
\begin{eqnarray}
ab=-n(\epsilon+\Gamma+\Delta+n-1) \nonumber \\
(1+\mid m \mid +\frac{\sqrt{1+E-i\gamma}-\sqrt{1+E+i\gamma}}{2})(1+\mid m \mid -\frac{\sqrt{1+E-i\gamma}+\sqrt{1+E+i\gamma}}{2})= \nonumber \\
-n(1-\sqrt{1+E+i\gamma}+2\mid m \mid +n+2) \nonumber \\
E_{n}=(n+\mid m \mid)(n+\mid m \mid+2)-\frac{\gamma^{2}}{4(n+\mid m \mid+1)^{2}}\nonumber\\
\end{eqnarray}

\begin{eqnarray}
ab=(n+\Delta+\epsilon)(\Gamma-n-1) \nonumber \\
(1+\mid m \mid +\frac{\sqrt{1+E-i\gamma}-\sqrt{1+E+i\gamma}}{2})(1+\mid m \mid -\frac{\sqrt{1+E-i\gamma}+\sqrt{1+E+i\gamma}}{2})= \nonumber \\(2\mid m \mid +n+2)(1-\sqrt{1+E+i\gamma}-n-1) \nonumber \\
E_{n}=(n+\mid m \mid)(n+\mid m \mid+2)-\frac{\gamma^{2}}{4(n+\mid m \mid+1)^{2}}\nonumber\\
\end{eqnarray}\\

\subsection{2. Two Coulombically repelling electrons on a sphere}
The problem of two electrons interacting via a Coulomb potential on a surface of a D-dimensional sphere of radius R is discussed in Ref.\cite{zhang} and the Heun equation is obtained for inter-electron wave function after introducing a dimensionless variable $z=u/2R$. ;
\begin{equation}
\label{k}
\frac{d^2\psi}{dz^2}+\bigg[\frac{\frac{1}{\gamma}}{z}+\frac{\frac{1}{2}\big(\delta-\frac{1}{\gamma}\big)}{z+1}+\frac{\frac{1}{2}\big(\delta-\frac{1}{\gamma}\big)}{z-1}\bigg]\frac{d\psi}{dz}+\frac{-4R^{2}E z+2R}{z(z+1)(z-1)}\psi=0.
\end{equation}
It is also shown that this equation has polynomial solutions of degree $n=1,2,...$;
\begin{equation}
\psi(z)=\prod\limits_{i=1}^{n}(z-z_{i}).
\end{equation}
stipulated that $E$ and $R$ take the values given by;
\begin{eqnarray}
E=\frac{n}{4R^{2}}(n+\delta -1)  \nonumber \\
R=-\frac{1}{2}[2(n-1)+\delta]\sum\limits_{i=1}^{n}z_{i}
\end{eqnarray}
where $z_{1}, z_{2},...z_{n}$ can be determined from Bethe ansatz equations \cite{zhang};
\begin{equation}
\sum\limits_{j\neq i}^{n}\frac{2}{z_{i}-z_{j}}+\frac{\frac{1}{\gamma}}{z_{i}}+\frac{\frac{1}{2}\big(\delta-\frac{1}{\gamma}\big)}{z_{i}+1}+\frac{\frac{1}{2}\big(\delta-\frac{1}{\gamma}\big)}{z_{i}-1}=0.
\end{equation}
Furthermore they calculated the radius $R$ and the energy $E$  as $R=\frac{1}{2}\sqrt{\frac{\delta}{\gamma}}$, $E=\gamma$ and $R=\frac{1}{2}\sqrt{2(\delta +2)+\frac{4\delta+6}{\gamma}}$, $E=\frac{\gamma(\delta +1)}{\gamma(\delta +2)+2\delta+3}$ for $n=1$ and $n=2$ states respectively.

This problem could be analysed by an easier and faster approach when extended NU method is used. When the parameters $\gamma, \delta, \epsilon, \alpha\beta, q,a$ in Eq.(\ref{eq:heun}) are replaced by $1/\gamma, 1/2(\delta-1/\gamma),1/2(\delta-1/\gamma) , -4R^{2}E, -2R,-1$ respectively, one can obtain  Eq.(\ref{k}). After this matching, the energy  $E$  and the radius $R$ (up to an integration constant) can be achieved  by writing new parameters in Eq.(\ref{heunI}) and Eq.(\ref{q1}) respectively:
\begin{eqnarray}
-4R^{2}E=-n(1/\gamma+1/2(\delta-1/\gamma)+1/2(\delta-1/\gamma)+n-1)  \nonumber \\
E=\frac{n}{4R^{2}}(n+\delta -1) \\
2R=C_{n}.
\end{eqnarray}
For the state $n=1$, the accessory parameter $q=2R$ must provide the determinant that admit polynomial solutions of Heun equation in Ref.\cite{ciftci}:

{\begin{tabular}{|cc|}
 $q$ & $-a\gamma$ \\ 
 $-\alpha\beta $ &  $q+a(\delta+\gamma)+\epsilon+\gamma$\\ 
\end{tabular} }=0.\\
For this problem this determinant has the following form:

{\begin{tabular}{|cc|}
 $C_{1}$ & $1/\gamma$ \\ 
 $\delta $ &  $C_{1}$\\ 
\end{tabular} }=0.\\
From this determinant, the constant $C_{1}$ can be obtained as follows;
\begin{equation}
C_{1}^{2}=\delta/\gamma.
\end{equation}
Thus the radius $R=\frac{1}{2}\sqrt{\frac{\delta}{\gamma}}$ (by choosing the positive root since $R$ is non-negative) and the energy $E=\gamma$ are obtained in agreement with those obtained in \cite{zhang}.

For $n=2$ state analogously to state of $n=1$, the accessory parameter $q=2R$ can be found from the solution of  a determinant:

{\begin{tabular}{|ccc|}
 $q$ & $-a\gamma$ & $0$ \\ 
 $-\alpha\beta $ &  $q+a(\delta+\gamma)+\epsilon+\gamma$ & $-2a(1+\gamma)$ \\ 
 $0$ & $-\alpha\beta-(\gamma+\epsilon+\delta)$ & $q+2(a+1)+2(a(\delta+\gamma)+\epsilon+\gamma)$ \\ 
\end{tabular} }=0\\
which is given in Ref.\cite{ciftci}. For the new parameters, this determinant is transformed into the following form;
 
{\begin{tabular}{|ccc|}
 $C_{2}$ & $1/\gamma$ & $0$ \\ 
 $2(1+\delta) $ & $C_{2}$ & $2(1+\gamma)$ \\ 
 $0$ & $2+\delta$ & $C_{2}$ \\ 
\end{tabular} }=0.\\
After solving for $C_{2}$, we get;
\begin{equation}
C_{2}^{2}=2(\delta +2)+\frac{4\delta+6}{\gamma}.
\end{equation}
Therefore the radius $R=\frac{1}{2}\sqrt{2(\delta +2)+\frac{4\delta+6}{\gamma}}$ and the energy $E=\frac{\gamma(\delta +1)}{\gamma(\delta +2)+2\delta+3}$ which are again in full agreement with those obtained in \cite{zhang} for the state $n=2$.

\subsection{3. Hyperbolic double-well potential}
The solution of Schr\"odinger equation for a particle of mass $m$ and energy $E$ in a potential $V(x)=-V_{0}\frac{sinh^{4}(z)}{cosh^{6}(z)}$ is discussed in Ref.\cite{downing}. One-dimensional Schr\"odinger equation for this particle is given by;
\begin{equation}
\frac{d^{2}}{dz^{2}}\psi(z)+\bigg(\epsilon d^{2}+U_{0}d^{2}\frac{sinh^{4}(z)}{cosh^{6}(z)}\bigg)\psi(z)=0
\end{equation}
where $\epsilon=2mE/\hbar^{2}$, $U_{0}=2mV_{0}/\hbar^{2}$ and $z=x/d$. In the aforesaid paper it is demonstrated that, this equation could be transformed into a confluent Heun equation;
\begin{equation}
\label{ch5}
\frac{d^{2}}{d\xi^{2}}y(\xi)+\bigg(\alpha+\frac{\beta +1}{\xi}+\frac{\gamma +1}{\xi-1}\bigg)\frac{d}{d\xi}y(\xi)+\bigg(\frac{\mu}{\xi}+\frac{\nu}{\xi-1}\bigg)y(\xi)=0
\end{equation}
where $\alpha=-d\sqrt{U_{0}}$, $\beta=-id\sqrt{\epsilon}$, $\gamma=-1/2$, $\mu=\frac{1}{4}\big(\alpha(\alpha+2)+2\alpha\beta-\beta(\beta-1)\big)$, $\nu=\frac{1}{4}\big(\alpha+\beta(\beta+1)\big)$.
To obtain the N-degree polynomial solutions of this confluent Heun equation, they have used the procedure which is reported in the first paragraph of Section 5 of this manuscript. After determining two termination conditions, the eigenvalue spectra for the symmetric $(s)$  and antisymmetric $(a)$ cases have been achieved as;
\begin{eqnarray}
\epsilon_{N}^{s}=-\frac{1}{4d^{2}}\bigg(3+4N-d\sqrt{U_{0}}\bigg)^{2}\\
\epsilon_{N}^{a}=-\frac{1}{4d^{2}}\bigg(5+4N-d\sqrt{U_{0}}\bigg)^{2}.
\end{eqnarray} 

This problem can be analysed by an easier and more efficient approach using the extended NU method. Since Eq.(\ref{ch5}) has the same form with  Eq.(\ref{che1}), the results which are obtained by extended NU method in Section 5, can be used directly for the eigenvalue spectra of this potential. Thus one can approach to the symmetric energy specta by using Eq.(\ref{sym1}) and Eq.(\ref{sym2}). When the parameters $\mu$ and $\nu$ defined for this problem are written in Eq.(\ref{sym1}), one obtains;
\begin{equation}
\label{ss}
\frac{\alpha}{4}(\alpha+3+2\beta)=-n\alpha.
\end{equation}
Subsituting $\alpha=-d\sqrt{U_{0}}$ and $\beta=-id\sqrt{\epsilon}$ in Eq.(\ref{ss}) the symmetric eigenvalue solution can be obtained exactly:
\begin{equation}
\epsilon_{N}^{s}=-\frac{1}{4d^{2}}\bigg(3+4N-d\sqrt{U_{0}}\bigg)^{2}.
\end{equation} 
This solution is also achieved from Eq.(\ref{sym2}). The parameter $\mu$ which was obtained in Eq.(\ref{mu1}) by extended NU method, can be modified for this problem for the state $N=1$;
\begin{equation}
\mu=\frac{1}{2}(2-\alpha+\beta+\gamma)+C_{12}.
\end{equation}
This result shows that the parameter $\mu$ can be determined up to an integration constant $C_{12}$. Using the $2^{nd}$ order determinant and the solution of related equation given in Section 5, one could make a verification of this result. From this solution which is obtained via root-finding methods, one could come to the conclusion: 
\begin{equation}
\mu=\frac{1}{2}(2-\alpha+\beta+\gamma)\pm\frac{1}{2}\sqrt{(2-\alpha+\beta+\gamma)^{2}-4\alpha(1+\beta)}.
\end{equation}
When the integration constant $C_{12}$ is equal to $\pm\frac{1}{2}\sqrt{(2-\alpha+\beta+\gamma)^{2}-4\alpha(1+\beta)}$, our solution will be the exact solution for parameter $\mu$.\\
Proceeding as before the antisymmetric energy specta is achieved from  Eq.(\ref{asym1}) or Eq.(\ref{asym2}). Ater substituting  parameters $\mu$ and $\nu$ into Eq.(\ref{asym1}), it can be easily shown that
\begin{equation}
\frac{\alpha}{4}(\alpha+3+2\beta)-\gamma \alpha=-n\alpha .
\end{equation}
This allows us to find the following antisymmetric eigenvalue spectra:
\begin{equation}
\epsilon_{N}^{a}=-\frac{1}{4d^{2}}\bigg(5+4N-d\sqrt{U_{0}}\bigg)^{2}.
\end{equation}
The same result can be obtained upon solving Eq.(\ref{asym2}) as well.
\section{VII. Conclusion}
Nikiforov-Uvarov (NU) method which is developed for solving second order differential equations by special orthogonal functions, is an easy and elegant method. This method has certain boundary comditions which are related to degrees of polynomial coefficients in the main equation of NU method. In turn, these boundary conditions bring some restrictions on the number of singular points of the relevant differential equations. Consequently, NU method will succeed only for equations which have at most three singular points. 

In this study, we propose that NU method is a method appicable for the equations which have at most four singularities. For this proposal the boundary conditions of NU method are changed and the extended NU method is obtained. Thereby, any second order differential equation which has at most four singularities, is reduced to the basic equation of extended NU method and analyzed systematically for eigenvalue and eigenfunction solutions. 

In this context, we have investigaed  Heun and confluent Heun equations which yield the boundary conditions of the extended NU method and obtained exact eigenvalue solutions without using traditional power series methods. The eigenvalue solutions of three physical problems have been worked out in order to demonstrate the accuracy of the extended NU method  and it has been observed that calculated energy eigenvalues are consistent with the former studies given in literature.

\bibliography{basename of .bib file}

\end{document}